\documentclass[reprint,aps,...]{revtex4-2}

\usepackage[a4paper, total={7in, 10in}]{geometry}
\usepackage[utf8]{inputenc}
\usepackage{amsmath}
\usepackage{amssymb}
\usepackage{bbm}
\usepackage{graphicx}
\usepackage[colorlinks, citecolor=blue]{hyperref}
\usepackage{xcolor}
\usepackage{multirow}
\usepackage{float}
\usepackage{placeins}
\usepackage{siunitx}
\usepackage{comment}

\def \p{\boldsymbol{p}}

\begin{document}

\title{Optimization of decoder priors for accurate quantum error correction}
\author{Volodymyr Sivak}
\email{vladsivak@google.com}
\author{Michael Newman}
\author{Paul Klimov}
\affiliation{Google Quantum AI}

\begin{abstract}
    Accurate decoding of quantum error-correcting codes is a crucial ingredient in protecting quantum information from decoherence.
    It requires characterizing the error channels corrupting the logical quantum state and providing this information as a prior to the decoder.
    We introduce a reinforcement learning inspired method for calibrating these priors that aims to minimize the logical error rate. Our method significantly improves the decoding accuracy in repetition and surface code memory experiments executed on Google's Sycamore processor, outperforming the leading  decoder-agnostic method by $16\%$ and $3.3\%$ respectively. This calibration approach will serve as an important tool for maximizing the performance of both near-term and future error-corrected quantum devices.
\end{abstract}

\maketitle

\section{Introduction}

Decoders are essential elements of the quantum error correction (QEC) stack \cite{dennis2002topological, fowler2012surface, bravyi2014efficient, higgott2023improved, higgott2023sparse, shutty2024efficient, wang2023transformer, bausch2023learning}. Given an observed error syndrome, they attempt to infer which underlying physical errors likely caused this syndrome. The decoder’s prior significantly impacts the accuracy of such an inference. Calibration of the prior is a well-known problem in classical error correction employed in wireless communication \cite{goldsmith2005wireless,ozdemir2007channel}. 
In the context of QEC, this problem gained relevance in light of the recent experimental advances with quantum devices based on superconducting circuits, trapped ions, and neutral atoms \cite{kelly2015state, ofek2016extending, google2021exponential, ryan2021realization, google2023suppressing, sivak2023real, bluvstein2024logical}.
Several approaches address this problem in a way that empirically improves the QEC performance under various assumptions \cite{spitz2018adaptive, google2021exponential, chen2022calibrated, google2023suppressing, wang2023dgr}. However, they generally optimize not for the actual metric of interest in QEC, the logical error rate (LER), but rather for various proxies of it, e.g., the mismatch between the observed syndrome statistics and the statistics predicted by the prior.

To address this vital limitation, we introduce a scalable method for calibrating the priors that directly aims to minimize the logical error rate of any given QEC decoder.
Relying on the assumption that physical errors are spatially localized, we use small-distance error correcting codes as local sensors of the error landscape in the device. The parameters of the error model are jointly optimized by a team of reinforcement learning agents \cite{sutton2018reinforcement} whose rewards are related to the decoding accuracy of the corresponding small sensor-codes. After the optimization, these parameters are used to weight an error hypergraph representing the prior for the larger target code that will be executed in the sensed region of the device. This decoder-aware optimization ensures the adaptation of the prior to both the true error channel of the device as well as the heuristics used by the decoder.

We demonstrate that our method improves upon several decoder-agnostic methods for calibrating the prior by benchmarking them on the QEC datasets acquired on Google's Sycamore quantum processor \cite{arute2019quantum}. 
By optimizing the error hypergraphs for a fast correlated matching decoder \cite{fowler2013optimal, higgott2023sparse}, we suppress the average LER of the distance-5 surface code by $10.6\%$ relative to an uninformative prior or $3.3\%$ relative to a correlation-based prior. For the distance-21 repetition code, the LER suppression is $48\%$ and $16\%$, respectively.
Our calibration method, which aims to directly minimize the logical error rate, will serve as an important tool for achieving near-optimal performance as we continue scaling the QEC technology.

\section{Small codes as error sensors \label{sec:sensor codes}}

In this work, we focus on the stabilizer formalism for QEC \cite{gottesman1997stabilizer}, in which the stabilizer operators are repeatedly measured to monitor the occurrence of errors. We define {\it detectors} as collections of stabilizer measurements that, in the absence of errors, will have a particular parity. During error correction, the {\it syndrome} indicates parity violations for any of the detectors. Within this framework, it is convenient to model the prior probability distribution of errors as a hypergraph in which nodes correspond to detectors and hyperedges connecting clusters of nodes correspond to various error mechanisms that activate those detectors \cite{gidney2021stim}. Every hyperedge is assigned an {\it a priori} probability of the corresponding error occurring in the device. 

\begin{figure}
    \centering
    \includegraphics[width=\columnwidth]{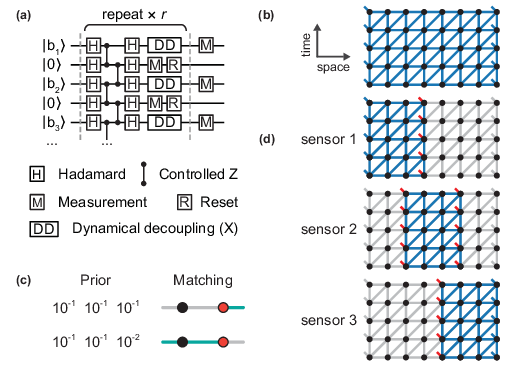}
    \caption{{\bf (a)} Error-correcting circuit for the repetition code memory. Data qubits are initialized with arbitrary bit strings $b_1b_2b_3...$, while auxiliary qubits are initialized in $|0\rangle$. The memory extends for $r$ cycles. {\bf (b)} Error hypergraph for the repetition code of distance $d=9$ and duration $r=4$ cycles under the assumption of one- and two-qubit Pauli noise channels applied to each circuit operation. It contains hyperedges of degree at most 2 (edges). The relation between Pauli errors in different locations of the circuit (a) and edges in (b) are elucidated in Ref.~\cite{google2021exponential}. {\bf (c)} Example of the MWPM decoding of a small error graph with two detectors where the syndrome indicates that a single detector (red circle) was triggered. The decoding outcome (green edges) changes depending on the prior (error probabilities for each edge). {\bf (d)} Example coverage of the target repetition code from (b) with three sensors of distance $d_s=5$ whose sensing regions (blue) overlap. A small fraction of edges (red) are present in the sensors but not in the target.}
    \label{fig1}
\end{figure}

The structure of the error hypergraph depends on the QEC circuit and on the choice of the modeled noise channels \cite{mcewen2023relaxing, cain2024correlated}. For clarity, in this section we illustrate our idea using the repetition code whose QEC circuit is shown in Fig.~\ref{fig1}(a). 
This circuit has a property that any one- or two-qubit Pauli error can trigger at most two detectors, and hence the resulting error hypergraph becomes a graph, shown in Fig.~\ref{fig1}(b). This graph-like nature of the prior allows the use of the minimum weight perfect matching (MWPM) algorithm to efficiently implement the  most-likely error decoder for the repetition code \cite{dennis2002topological,fowler2012surface}. 
As illustrated in Fig.~\ref{fig1}(c), the prior probability assignment is important because it directly affects the decoding outcome of MWPM for a given observed error syndrome.

For a repetition code of distance $d$ and duration $r$ cycles, the number of edges in the error graph in Fig.~\ref{fig1}(b) is $3r(d-1)+d$. Although nominally independent, some of these edges correspond to the same physical error mechanism that is replicated in time. Hence, it is reasonable to use time-translation symmetry to reduce the size of the error graph parametrization by identifying error equivalence classes based on this symmetry. In real devices, additional error sources such as the accumulation of leakage, can lead to non-stationary QEC processes \cite{mcewen2021removing, battistel2021hardware, miao2023overcoming}. However, we can still impose the time-translation symmetry on our prior -- a choice we will exploit to keep the size of the parametrization relatively small. In the case of the repetition code, it leads to $9(d-1)$ independent parameters: for the bulk and each of the two time boundaries we parameterize $d$ space-like, $d-1$ time-like and $d-2$ spacetime-like edges \cite{google2021exponential}. 

To achieve the best performance, it is preferable to learn the values of the $9(d-1)$ parameters describing the prior by minimizing the logical error rate of the code. However, unlike the decoder-agnostic methods \cite{spitz2018adaptive, google2021exponential, chen2022calibrated, google2023suppressing}, this strategy immediately faces the challenge of exponential sample complexity required to even resolve the LER, since it scales as ${\rm LER}\propto \Lambda^{-d/2}$, where $\Lambda$ is a constant characterizing the device's error suppression capability \cite{fowler2012surface}. For repetition codes, $\Lambda\approx3$ has already been achieved with superconducting circuits \cite{google2021exponential}. A similar sample complexity challenge is encountered by the machine learning decoders trained to minimize the LER \cite{varbanov2023neural, lange2023data, wang2023transformer, bausch2023learning}. Despite this challenge, brute-force optimization could be used for small problem instances, as was shown with the Nelder-Mead algorithm in Ref.~\cite{kelly2015state}. 

To address this challenge in a scalable way, we use small error-correcting codes as local sensors of the error hypergraph. Such small codes can be executed before the target experiment to probe the error landscape in the device and calibrate the decoder prior. To illustrate this concept, consider the example in Fig.~\ref{fig1}(d) where three sensors of distance $d_s=5$ fully cover the $d=~9$ target repetition code. Each sensor shares most of its error mechanisms with the target code (and some with other sensors), but also introduces additional error mechanisms that do not affect the target code. Hence, the size of the common parametrization shared among all the sensors is larger than for the target code alone. For example, in Fig.~\ref{fig1}(d) it equals $9(d-1)+12$, where the additional $12$ parameters are related to the boundaries of the sensors. 

To learn the sensor error graph parameters, we now only require enough data to resolve the LER of the sensors. When estimating the LER with constant relative precision, the number of QEC shots scales as $O(d\Lambda^{d_s/2})$ for the repetition code and $O(d^2\Lambda^{d_s/2})$ for the surface code, where the polynomial dependence on $d$ comes from the number of sensors. Hence, by using fixed-size sensors we avoid the exponential overhead. 

\section{Learning sensor parameters}

\begin{figure}
    \centering
    \includegraphics[width=\columnwidth]{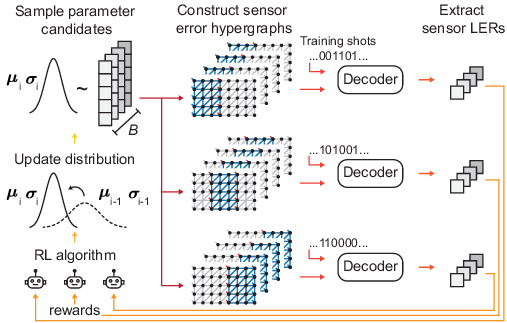}
    \caption{Multi-agent reinforcement learning pipeline for training error hypergraphs. The training epoch $i$ begins with sampling a batch of $B$ parameter candidates from the policy. The sampled parameters are used to construct error hypergraph candidates for each sensor, which are then evaluated on a training dataset of QEC shots to estimate the LERs. A distinct learning agent is associated to every sensor, and their rewards are assigned as $-\log_{10}({\rm LER})$. The gradients for the policy distribution are computed from the information available to all agents. Finally, the policy is updated and the next training epoch begins.}
    \label{fig2}
\end{figure}

Having established a common parametrization of the hypergraph shared among all the sensors, we now need to optimize it towards multiple objectives related to the LERs of the sensors.

The specifics of this problem necessitate optimization algorithms compatible with high dimensionality (the number of parameters scales as $O(d^2)$ in the surface code, and already reaches $\sim10^3$ for $d=5$), stochasticity (LER estimation is subject to sampling noise) and generalization (the optimized error hypergraphs must be extrapolated to quantum memories of arbitrary duration). 

Problems with similar properties are encountered in reinforcement learning \cite{sutton2018reinforcement, mnih2015human, schulman2017proximal}, where large neural networks are optimized using the reward signal whose gradient is estimated with Monte Carlo techniques \cite{mohamed2020monte}. Here, we use an optimization approach inspired by multi-agent reinforcement learning (RL), where a team of learning agents collaborates to achieve a common goal, as illustrated in Fig.~\ref{fig2}. 

In this learning framework, we associate a Gaussian distribution to each optimizable parameter from the common parametrization. The resulting factorized multivariate distribution, called a {\it policy}, is characterized by the mean vector $\boldsymbol{\mu}$, which represents the current best guess of the optimal error model parameters, and a diagonal covariance matrix $\boldsymbol{\sigma}^2$. Over the course of multiple training epochs, the learning agents jointly reshape this policy, gradually shifting its mean towards the locally optimal solution and shrinking its variance to localize the solution more accurately. 

Each agent is responsible for only part of the policy related to a particular sensor. It affects that part of the policy by applying small updates to its mean and covariance matrix along the estimated gradient of this agent's reward function. 
Since in general the sensors can spatially overlap, as exemplified in Fig.~\ref{fig1}(d) and further elaborated in Appendix~\ref{appendix:rep_code_variants}, some parts of the policy could be affected by several agents simultaneously. 

Fig.~\ref{fig2} shows how we produce the rewards for the learning agents. First, we sample a batch of parameter candidates from the current policy.  Then, we construct the error hypergraph for each sensor using the relevant part of the sampled common parametrization. Next, we decode the training shots of each sensor to estimate its LER. Finally, we assign the rewards to the corresponding agents as $-\log_{10}({\rm LER})$. The Monte Carlo gradients \cite{mohamed2020monte} are computed from the rewards following a generalization of the proximal policy optimization (PPO) algorithm \cite{schulman2017proximal} to the multi-agent setting, see Appendix~\ref{appendix:RL}.

\section{Results}

\begin{figure}
    \centering
    \includegraphics[width=\columnwidth]{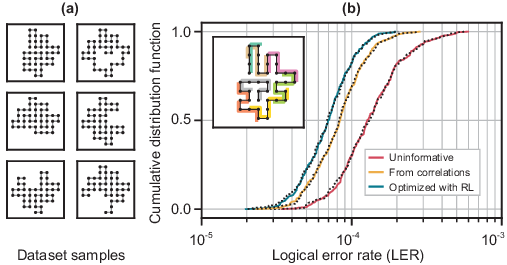}
    \caption{{\bf (a)} Several samples from a dataset of $d=21$ repetition codes. Dots represent qubits in the device, and lines show the repetition code connectivity. {\bf (b)} Integrated histogram of the LERs of all dataset samples obtained for a MWPM decoder and different methods of generating the prior (color). The performance on the test set (solid lines) is similar to the training set (dotted black lines) indicating the absence of overfitting. Inset shows how one particular $d=21$ repetition code is covered by overlapping $d_s=5$ sensors.}
    \label{fig3}
\end{figure}

We first benchmark our method on experimental data from $d=21$, $r=21$ repetition codes. The dataset contains $266$ code instances laid out in random configurations on a square grid of qubits, with a few representative samples shown in Fig.~\ref{fig3}(a). Each sample contains $2\times 10^5$  shots, which we split into $5\times 10^4$ for training and validation, and $1.5\times 10^5$ for testing. 

When mapped onto a real device, most gates in the repetition code circuit are of high fidelity. However, certain gates suffer from enhanced leakage, crosstalk, energy decay or other imperfections \cite{klimov2024optimizing}. As a result, the range of LERs among the dataset samples covers more than an order of magnitude. This allows us to asses the relative advantages of various decoding priors and methods under a diverse set of conditions and to draw statistically significant conclusions.

In Fig.~\ref{fig3}(b), we compare the LER obtained with a MWPM decoder configured with three different priors: (i) an uninformative prior, (ii) a prior obtained by fitting the pairwise correlations in the error graph to the correlations observed in the data \cite{spitz2018adaptive,google2021exponential,chen2022calibrated,google2023suppressing}, and (iii) a prior obtained via RL with $d_s=5$ sensors (in this case, the sensors are subsampled from the target code \cite{google2021exponential}). We give the methods (ii) and (iii) access to the same set of training data, and use the test set to compare the final performance. More information on methods (i) and (ii) is provided in Appendix~\ref{appendix:alternative_priors}.

The benchmarking results indicate that calibrating the prior leads to $48\%$ average reduction of LER compared to an uninformative prior (i). Moreover, our RL-based method outperforms the correlation-based method (ii) by $16\%$ on average, and this result is largely independent of the number of sensors used for training (7, 9, or 17) and the method used to seed the initial parameter values [(i) or (ii)], see Appendix~\ref{appendix:rep_code_variants} for details. 

Our analysis so far was restricted to the repetition code with one- and two-qubit Pauli noise channels, which has a graph-like error model. However, in general the error models of the QEC codes are represented as hypergraphs. 
For example, the surface code under the same noise channels contains hyperedges of degree four in the bulk. 
While some decoders can operate with the hypergraph error model directly \cite{bravyi2014efficient, cain2024correlated}, others decompose the hyperedges of the error model into sets of edges whose associated probabilities are derived from the probability and structure of the hyperedge \cite{higgott2023sparse, shutty2024efficient}. This decomposition is based on various empirically justified heuristics \cite{shutty2024efficient,wang2023dgr}.
RL can implicitly exploit these heuristics to tailor the prior to the specific decoder used during the training, which is a particular advantage of our method. 

\begin{figure}
    \centering
    \includegraphics[width=\columnwidth]{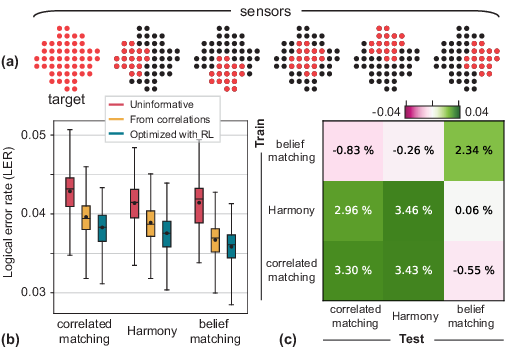}
    \caption{{\bf (a)} Coverage of the $d=5$ target surface code with $d_s=3,5$ sensor codes. During the optimization we use $d=5$ code as an additional sensor (see main text). {\bf (b)} Logical error rates of the target $d=5$ code obtained with different decoders and different priors. The box extends from the first to the third quartile of the data, with a line at the median and a circle at the mean. The whiskers indicate the full range of the data.  {\bf (c)} Average relative improvement of the LER when using RL as compared to the decoder-agnostic method (ii) used to seed the training. Negative values indicate that the prior learned with a specific decoder (marked ``Train'') does not generalize to another decoder (marked ``Test''). }
    \label{fig4}
\end{figure}

To benchmark our method on the general hypergraph error models, we use a dataset of $d=3,5$ surface codes (the XZZX variant \cite{bonilla2021xzzx}) of varying duration $r\in(5, 9, 13, 17, 21, 25)$ prepared and measured in the $X$ basis. The dataset consists of 53 experiments, each with a single $d=5$ patch and five $d=3$ patches, see Fig.~\ref{fig4}(a). There are $7.5\times10^4$ shots for every $(d,r)$ combination. 
We also note that the Sycamore device on which this data was acquired does not reach the $\Lambda\approx1$ performance that was previously demonstrated by Google \cite{google2023suppressing}.

As we show in Appendix~\ref{appendix:sensor_size_surf}, distance-3 sensors are insufficient to fully cover larger target codes without the omission of any hyperedges. Hence, we made the choice to include the target $d=5$ patch itself as a sensor {\it in addition} to five small $d_s=3$ sensors. 
This choice results in a 964-dimensional common parametrization in which 748 parameters belong to the $d = 5$ target code. We also note that distance-5 sensors are sufficient for any larger target code.

Utilizing the time invariance of our parametrization, we construct hypergraphs for any code duration $r$ using the values of parameters learned at duration $r=5$. Since the data from longer-duration experiments is not used for training, this procedure automatically provides a train-test separation and demonstrates the generalization of the learned error hypergraphs in a practically-relevant scenario. We fit the decay of the logical observable to an exponential to extract the LER as detailed in \cite{google2023suppressing} and exclude the $r=5$ training datapoint from this fit.  

We compare the same three methods of producing the prior as in Fig.~\ref{fig3}(b) and three different decoders: correlated matching \cite{higgott2023sparse}, Harmony \cite{shutty2024efficient}, and belief-matching  \cite{higgott2023improved}, with results summarized in Fig.~\ref{fig4}(b). We find that a combination of the belief-matching decoder with optimized priors attains the best performance in this above-threshold error regime ($\Lambda<1$). 

To demonstrate that RL indeed exploits the decoder heuristics to maximize the performance gain from the prior, we compare  the LERs obtained by running the decoder with a prior trained for another decoder, see Fig.~\ref{fig4}(c). In most cases, the prior does not fully generalize across decoders. For example, if the prior is trained for the correlated matching decoder and then used with the belief-matching decoder, the performance would be on average $0.8\%$ worse than using a decoder-agnostic prior from method (ii). However, we find strong generalization among correlated matching and Harmony, which is expected since Harmony is based on ensembling of correlated matching decoders.

\section{Conclusion}

In conclusion, we demonstrated that calibration of the prior is a necessary ingredient towards high-accuracy decoding in QEC. Our calibration method inspired by multi-agent reinforcement learning outperforms the studied alternative approaches and possesses a lot of desirable properties: it is highly parallelizable, scalable to large code distances, compatible with arbitrary decoders and with arbitrary hypergraph structures of the prior. 

Nevertheless, many challenges remain and present opportunities for further research. Despite the favorable asymptotic scaling estimates, the run time of our approach must be significantly improved for compatibility with real-time operation of the device. In particular, reinforcement learning is known to have relatively low sample efficiency \cite{sutton2018reinforcement}, and other algorithms might prove useful for faster joint optimization of the sensor parameters. In addition, it remains an open question how to systematically achieve generalization of the learned priors from fast and less accurate to slow but more accurate decoders. Finally, our optimization is performed on a fixed hypergraph structure, and we expect potential gains from adapting this structure to the data. This might more accurately account for realistic device imperfections such as crosstalk and leakage.

To facilitate further research into these questions and to enable practically-relevant benchmarking of the decoding innovations, we made the datasets used in this work publicly available in Ref.~\cite{dataset_sycamore_2024}.

\section{Acknowledgements}

We thank A. Bourassa for software assistance, J. Atalaya for prior work on the priors, J. Bausch, F. Heras, A. Eickbusch, O. Higgott, N. Shutty, S. Boixo, A. Morvan, and M. McEwen for feedback on the manuscript, and the entire Google Quantum AI team for maintaining the hardware, software, cryogenics, and electronics infrastructure that enabled this work.

\appendix

\section{Time-invariant parametrization\label{appendix:time_invar_param}}

\begin{figure*}
    \centering
    \includegraphics[width=\textwidth]{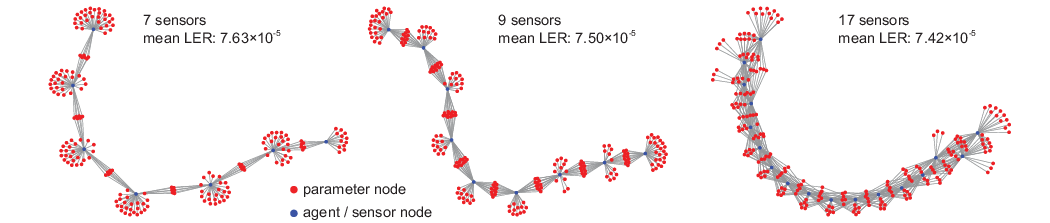}
    \caption{Graphical representation of the optimization problem for $d=21$ repetition code covered with $d_s=5$ sensors. Red nodes represent the parameters of the common parametrization of the error hypergraphs for all the sensors. Blue nodes represent the learning agents associated with the sensors. The graph connectivity indicates which parameters influence the rewards of which agents. Equivalently, it shows which agents control which parameters through Monte Carlo gradients, as described in Appendix~\ref{appendix:RL}. Increasing the number of sensors leads to enhanced parameter sharing among the agents, with an average degree of the parameter nodes being $1.2$, $1.4$, and $2.6$ when the number of agents is $7$, $9$ and $17$ respectively. The LER improvement of the target code from increasing the number of sensors hits diminishing returns, given the reference average LER of $8.97\times10^{-5}$ for the correlation-based prior using method (ii).}
    \label{fig:factor_graphs}
\end{figure*}

In this work, we use the following time-invariant parametrization of the error hypergraphs. Detectors are defined as sets of stabilizer measurement outcomes $\{m_1, m_2, ...\}$ whose total parity in the error-free circuit is deterministic. Each measurement $m_i$ can be characterized by its  space-time location $(x_i, y_i, t_i)$ in the circuit. We assign detector coordinates as sets of space-time coordinates of all measurements involved in this detector $\{(x_1, y_1, t_1), (x_2, y_2, t_2), ...\}$. Error mechanisms, represented as hyperedges, are defined by the sets of detectors that they trigger. We assign hyperedge coordinates as sets of coordinates of all detectors connected by the hyperedge, i.e. sets of sets of space-time coordinates of the measurements: 
\begin{align}
    \big\{ 
        & \{(x_1, y_1, t_1), (x_2, y_2, t_2), ...\}, \nonumber \\ 
        & \{(x_1', y_1', t_1'), (x_2', y_2', t_2'), ...\}, \nonumber 
        ...\big\}
\end{align}
Note that the size of the ``coordinate vector" can vary depending on the number of detectors and measurements involved in the hyperedge. 

We use time-translation symmetry to define equivalence classes: hyperedges whose coordinates differ by translation $\delta t$ in time 
\begin{align}
    \big\{ 
        & \{(x_1, y_1, t_1+\delta t), (x_2, y_2, t_2+\delta t), ...\}, \nonumber \\ 
        & \{(x_1', y_1', t_1'+\delta t), (x_2', y_2', t_2'+\delta t), ...\}, \nonumber 
        ...\big\}
\end{align}
belong to the same class and are assigned the same parameter value.

For circuits that contain multiple repetitions of identical QEC cycles, this construction properly identifies the equivalent error mechanisms in the bulk and distinguishes them from the error mechanisms at time boundaries, where the definition of detectors relies either on the final measurements of the data qubits in the last cycle or the initially prepared states of the data qubits before the first cycle. 
  
\begin{figure*}
    \centering
    \includegraphics[width=\textwidth]{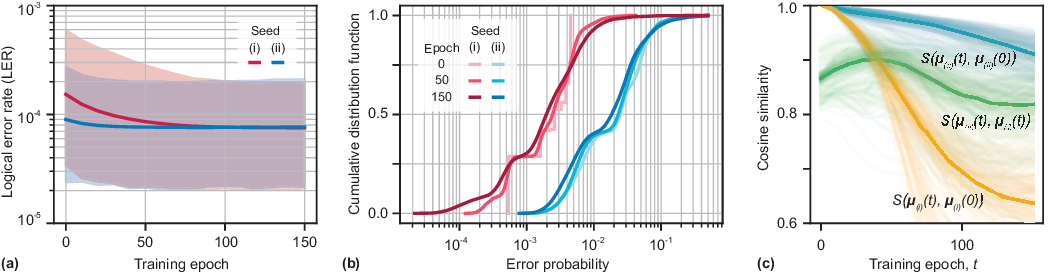}
    \caption{{\bf (a)} Convergence of the distribution of the logical error rates in a repetition code dataset starting from different initial seeds, where the initial mean $\boldsymbol{\mu}$ of the Gaussian policy is set using either method (i) or (ii). The solid lines show the mean of the LER distribution across all dataset samples, and the shaded region extends from the minimal to the maximal LER. {\bf (b)} Cumulative distribution function of all hyperedge probabilities across all dataset samples, taken at different training epochs and for different initial seeds. The deformation of the distribution due to the training is more significant when starting from an uninformative prior (i). {\bf (c)} Evolution of the cosine similarity of hypergraph parametrizations (defined in Appendix~\ref{appendix:rep_code_variants}) during the training. Lightly colored lines correspond to different samples from the repetition code dataset, and solid lines show the mean across all samples.}
    \label{fig:training_convergence}
\end{figure*}

\section{Repetition code training variations \label{appendix:rep_code_variants}}

We experimented with different training variations for the repetition code dataset to understand the robustness of the RL approach. 

As can be seen from the graphical representation of the optimization problem in Fig.~\ref{fig:factor_graphs}, increasing the number of sensors by making them overlap on more qubits leads to enhanced parameter sharing. In principle, this could be beneficial, as it would ensure that joint parameter optimization does not overfit to any single sensor. However, we find only marginal performance gains from this increase of the sensor count: the average LER reduces from $7.63\times 10^{-5}$ to $7.42\times 10^{-5}$ when the number of sensors is more than doubled. Hence, sensor coverage with minimal overlap might be preferable from a practical standpoint to minimize the  training time. 

In addition, we compare the trainings starting from different initial parameter values, using either method (i) or (ii) to seed the mean $\boldsymbol{\mu}$ of the Gaussian policy and keeping all other hyperparameters the same. As shown in Fig.~\ref{fig:training_convergence}(a), we find that regardless of the seed the training converges to similar LERs, although the seed has a significant effect on the convergence speed. It takes about 100 epochs to converge when starting from an uninformative prior (i) and only about 30 epochs when starting from a correlation-based prior (ii).

Although the final distribution of LERs is independent of the initial seed, the hyperedge probabilities to which the optimization converges differ significantly. The main difference is in the absolute scale of the hyperedge probabilities, as seen in Fig.~\ref{fig:training_convergence}(b) where the probabilities learned using seed (i) are overall about an order of magnitude smaller than using seed (ii). Although seemingly large, this discrepancy is irrelevant, because scaling all probabilities by the same factor does not affect the decoding outcome of the MWPM decoder. However, the shapes of the final hyperedge distributions are also different, implying that optimization does not converge to equivalent local minima. 

To quantify the discrepancy of the parameters learned starting from different seeds, we measure the cosine similarity between the means of the Gaussian policies during the optimization process. This similarity measure, defined as $S(\boldsymbol{\mu}_1,\boldsymbol{\mu}_2)=\boldsymbol{\mu}_1/|\boldsymbol{\mu}_1| \cdot \boldsymbol{\mu}_2/|\boldsymbol{\mu}_2|$, is agnostic to the global rescaling of the parameters and hence it is particularly well-suited for our problem. As shown in Fig.~\ref{fig:training_convergence}(c), the similarity $S\big(\boldsymbol{\mu}_{(i)}(t),\boldsymbol{\mu}_{(i)}(0)\big)$ reduces during the training, which simply means that the learned parameters progressively deviate away from the initial seed. The similarity $S\big(\boldsymbol{\mu}_{(ii)}(t),\boldsymbol{\mu}_{(ii)}(0)\big)$ also reduces but to a smaller extent, since seed (ii) is closer to the optimal solution. Finally, the similarity $S\big(\boldsymbol{\mu}_{(ii)}(t),\boldsymbol{\mu}_{(i)}(t)\big)$ first increases, indicating that optimization deforms the uninformative prior in the direction of the correlation-based prior, and then reduces, indicating that the final local minima found from different seeds are not equivalent.

\begin{figure*}
    \centering
    \includegraphics[width=\textwidth]{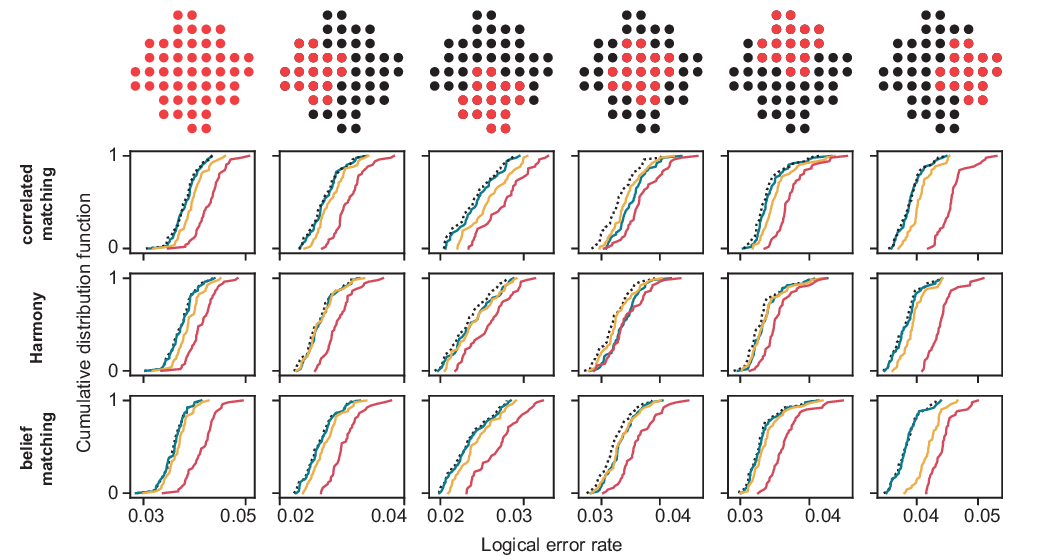}
    \caption{Cumulative distribution functions of LERs in the surface code dataset for each sensor. The top panel shows (in red) which qubits in the grid belong to which sensor. In the histograms, red lines correspond to uninformative prior (i), yellow lines correspond to the correlation-based prior (ii), and green lines correspond to the RL prior (iii) with parameter sharing among all sensors. Black dotted lines show the performance of RL achieved without parameter sharing. Relaxing this constraint leads to slightly enhanced performance for all patches, especially pronounced in the middle $d_s = 3$ sensor.}
    \label{fig:per_sensor_cdf_surface_code}
\end{figure*}

\section{Alternative constructions of the prior \label{appendix:alternative_priors}}

In method (i), referred to as uninformative prior, we start with a noiseless Clifford circuit in Stim \cite{gidney2021stim} and append single-qubit depolarizing channels to all circuit moments with single-qubit operations, and two-qubit depolarizing channels to all circuit moments with CZ gates. The magnitude of all depolarizing channels is set to $10^{-3}$ regardless of the type of the operation. The error hypergraph is constructed from this noisy circuit by tracking which detectors are affected by each inserted Pauli error. The resulting prior does not have uniform hyperedge probabilities because certain clusters of detectors can be triggered by multiple combinations of Pauli errors inserted into the circuit. Nevertheless, the distribution of hyperedge probabilities has support on only a few distinct values, as can be seen in Fig.~\ref{fig:training_convergence}(b) for seed (i) and epoch 0.

In the correlation-based method (ii), we use a subset of the QEC shots to fit the prior ($5\times10^{4}$ shots in the repetition code dataset, and $7.5\times10^4$ shots from $r=5$ cycles experiment in the surface code dataset). We choose the structure of the error hypergraph to be identical to the prior (i), but update the hyperedge probabilities to ensure that the statistics of the syndromes sampled from the resulting hypergraph matches the statistics observed in the data. To achieve this, we match the  correlation functions of the data to the correlation functions of the hypergraph using an analytic approach described in detail in Refs.~\cite{spitz2018adaptive, google2021exponential, google2023suppressing}. This is done for up to four-body correlations in the surface code and two-body correlations in the repetition code.
When the statistics of the syndrome data cannot be faithfully captured by the imposed hypergraph structure, the values of certain hyperedges can become nonphysically small or even negative \cite{google2021exponential, google2023suppressing, chen2022calibrated}. In this case, instead of discarding the hyperedges, we floor them to empirically chosen thresholds, $10^{-2}$ for degree-1 hyperedges and $10^{-5}$ for higher-degree hyperedges.

\section{Minimal size of the surface code sensors \label{appendix:sensor_size_surf}}

\begin{figure}
    \centering
    \includegraphics[width=\columnwidth]{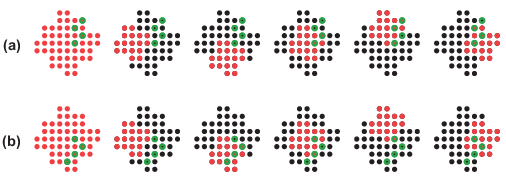}
    \caption{Space-like projections of the hyperedge coordinates for two examples from Appendix~\ref{appendix:sensor_size_surf}. These projections are obtained by highlighting the qubits whose measurements are involved in the detectors connected by the hyperedge. Hyperedges in both (a) and (b) connect clusters of four detectors. This shows that distance-3 sensors are insufficient for capturing all possible hyperedges in a surface code with one- and two-qubit Pauli noise channels.}
    \label{fig:min_sensor_size}
\end{figure}

Following the definition of the time-invariant hyperedge coordinates from Appendix~\ref{appendix:time_invar_param}, we find that certain classes of errors cannot be captured by the distance-3 sensors. For example, a hyperedge with coordinates
\begin{align}
    \big\{ 
        & \{(7, 7, 0), (7, 7, 1)\}, 
        \{(8, 6, 0), (8, 6, 1)\}, \nonumber \\ 
        & \{(8, 8, 0), (8, 8, 1)\}, 
        \{(7, 5, 0), (7, 5, 1)\} \nonumber 
        \big\},
\end{align}
or a hyperedge corresponding to a hook error
\begin{align}
    \big\{ 
        & \{(7, 3, 0), (7, 3, 1)\}, 
        \{(8, 4, 1), (8, 4, 2)\}, \nonumber \\ 
        & \{(6, 2, 0), (6, 2, 1)\}, 
        \{(7, 5, 0), (7, 5, 1)\} \nonumber 
        \big\},
\end{align}
whose space-like projections are shown in Fig.~\ref{fig:min_sensor_size} (a) and (b) respectively, can only be captured in our system by including the $d=5$ patch itself as an additional sensor. We find numerically that $d_s=5$ sensors are sufficient to capture all the error mechanisms of odd-distance target codes with $d\geq5$. 

We note also that our choice of parametrization, which is based solely on the space-time coordinates of the stabilizer measurements and ignores the pattern of gates in between those measurements, can lead to suboptimal sensing of the hyperedges that fall at the boundaries of the sensors. For example, a time-like error on a weight-2 stabilizer in a sensor would belong to the same equivalence class as its weight-4 counterpart from the bulk of the target code. 
Although this constraint does not seem to impair the performance, it can be mitigated by using denser sensor coverage, sensors of larger size, or a more elaborate choice of parametrization.

\section{LER of the surface code sensors \label{appendix:surf_code_histograms}}

By inspecting the LER distributions of the individual sensors in the surface code dataset, shown in Fig.~\ref{fig:per_sensor_cdf_surface_code}, we find that in addition to the improvement of the $d=5$ target code demonstrated in Fig.~\ref{fig4}(b), the $d_s=3$ sensor codes are also systematically improved. The central sensor, whose performance is degraded after the training, is the only clear exception from this general trend. Since this sensor shares the largest fraction of its parameters with the boundaries of the surrounding sensors, we conjecture that this sacrifice was necessary for RL to achieve the highest overall performance.

We consider an alternative training scenario, where the error hypergraph parameters are optimized for each surface code patch individually, without enforcing the constraint that parameters must be shared by all sensors. In this case there are no tradeoffs leading to the sacrifice of the center sensor performance. To confirm this and to evaluate the overall performance loss from imposing the parameter sharing constraint, we carried out such an alternative training, the results of which are shown in Fig.~\ref{fig:per_sensor_cdf_surface_code} with dotted black lines. As expected, the performance is slightly enhanced compared to the parameter-sharing case for most patches. However, as explained in Section~\ref{sec:sensor codes}, such direct training approach would become infeasible at large code distances.

\section{Configuration of the decoders \label{appendix:decoder_config}}

Our correlated matching decoder is based on the sparse blossom matching engine \cite{higgott2023sparse} and uses a variant of the two-step re-weighting strategy \cite{fowler2013optimal}. Our belief matching decoder \cite{higgott2023improved} is based on the same matching engine and has 10 belief propagation steps. The Harmony decoder uses hyperparameters from \cite{shutty2024efficient}. Its ensemble size is 11 during the training and 21 during the final decoding to get the results reported in Fig.~\ref{fig4}.

\section{Optimization with RL \label{appendix:RL}}

In this section, we provide a pedagogical derivation of our optimization algorithm. Many aspects of the derivation follow established techniques from the RL literature. However, to the best of our knowledge, the variance reduction method based on the sparsity of the agent-to-parameter mapping has not been presented in this form before.

\subsection{Optimization problem}

Consider the following optimization problem:
\begin{align}
    \p_{\rm opt} = \underset{\p}{\rm argmax}\,[R(\p)], \label{opt_problem}
\end{align}
where $\p\in \mathbb{R}^P$ is the parameter vector, and the function $R(\p): \mathbb{R}^P\to \mathbb{R}$ is a sum of $A$ local terms that depend only on the subsets of components of vector $\p$:
\begin{align}
    R(\p) = \sum_{a=1}^{A} R_a(\p_{a}),
\end{align}
where we use the notation $\p_a$ to denote the set of components of vector $\p$ that are involved in the function $R_a$. We use a different notation $p_i\equiv \p[i]$ to denote the $i$'s component of vector $\p$. 

In our error-hypergraph optimization problem, the functions $R_a$ correspond to $-\log_{10}({\rm LER}_a)$ for each sensor $a$. For training the $d=21$ repetition code in Fig.~\ref{fig3} we use $A=7$ and $P=198$, and for training the $d=5$ surface code in Fig.~\ref{fig4} we use $A=6$ and $P=964$. 

\subsection{Reinforcement learning}

Now let us cast the optimization problem in Eq.~\eqref{opt_problem} in the language of reinforcement learning. Consider a parametrized probability distribution $\pi_\theta(\p)$ called the {\it policy}, where $\theta$ denotes all variables parametrizing this policy. In this work, we used a simple Gaussian policy
\begin{align}
    \pi_\theta(\p) = \frac{\exp\left(-\frac{1}{2}(\p-\boldsymbol{\mu})^T\boldsymbol{\Sigma}^{-1}(\p-\boldsymbol{\mu})\right)}{\sqrt{(2\pi)^P\det(\boldsymbol{\Sigma}) }},  \label{gaussian_policy}
\end{align}
with diagonal covariance matrix $\boldsymbol{\Sigma}={\rm diag}[\boldsymbol{\sigma}^2]$. The parameters of this policy are $\theta = (\boldsymbol{\mu}, \boldsymbol{\sigma})$; in more complex cases, $\theta$ would include the parameters of a neural network parametrizing a deep policy \cite{mnih2015human}. 

The performance measure $J(\theta)$ of a policy $\pi_\theta(\p)$ is the expected value of a stochastic variable $R(\p)$ where parameter vectors $\p$ are sampled from this policy: 
\begin{align}
    J(\theta) = \underset{\p\sim \pi_\theta}{\mathbb{E}}[R(\p)]. \label{grad_define}
\end{align}

We can reformulate the optimization problem stated in Eq.~\eqref{opt_problem} as a problem of finding the optimal policy with respect to the performance measure $J(\theta)$. These problems are equivalent, since the optimal policy, when it is unique, is a deterministic policy that returns $\p_{\rm opt}$: 
\begin{align}
    \pi_{\rm opt}(\p) = \delta(\p - \p_{\rm opt}).
\end{align}

In policy-gradient reinforcement learning \cite{sutton2018reinforcement}, to find the optimal policy starting from an arbitrary initial policy $\pi_\theta(\p)$, we follow the gradient of the performance measure $\nabla_\theta J(\theta)$ and iteratively update $\theta$ until convergence. 
In practice, to obtain the gradient we sample a batch of candidate vectors $\p^{(1)}, ..., \p^{(B)}$ from the policy $\pi_\theta(\p)$, and then evaluate all the terms $R_a$ (not only the global $R$) for these candidates. This information is then used to construct an unbiased Monte Carlo estimator of $\nabla_\theta J(\theta)$ \cite{mohamed2020monte}. In the RL framework, the values of $R_a(\p)$ play the role of rewards of $A$ learning agents who jointly optimize the shared policy $\pi_\theta(\p)$.

\subsection{Derivation of the gradient estimator}

Next, we describe the construction of the gradient estimator \cite{mohamed2020monte}. There are numerous options for the unbiased estimator of the gradient (converging to the true gradient with an infinite sample size from the policy). To facilitate efficient learning, an estimator with minimal variance is required.  

To derive such a low-variance estimator, we begin by transforming the gradient of Eq.~\eqref{grad_define}:
\begin{align}
    \nabla_\theta J(\theta) 
    & = \nabla_\theta \underset{\p\sim \pi_\theta}{\mathbb{E}}[R(\p)] \label{gradient_of_J}
    \\ & = \nabla_\theta \int R(\p)  \pi_\theta (\p) d\p
    \\ & = \int R(\p) \nabla_\theta \pi_\theta (\p) d\p \label{grad_integral}.
\end{align}

Keeping the estimator unbiased, we can subtract an arbitrary {\it baseline} $b$ from $R(\p)$ in Eq.~\eqref{grad_integral}, as long as this baseline does not depend on $\p$. This is the case because $\int b\,  \nabla_\theta \pi_\theta (\p) d\p=b\,  \nabla_\theta \int \pi_\theta (\p) d\p = b  \nabla_\theta 1 = 0$. Note that the baseline $b$ can be arbitrary, even a random variable, and it can depend on the policy parameters $\theta$.

Intuitively, the optimal baseline approximately corresponds to the expected reward under the current policy. We will discuss how to choose $b$ in more detail shortly, but for now we impose that $b$ has a similar decomposition to $R$, i.e. it is a sum of $A$ terms corresponding to each agent: 
\begin{align}
    b = \sum_{a=1}^A b_a.
\end{align}

Next, we define the {\it advantage} functions 
\begin{align}
    \alpha_a(\p_a)=R_a(\p_a) - b_a.
    \label{advantage}
\end{align}
They are such that $\p$-candidates with rewards larger than the baseline have a positive advantage, while those with reward smaller than the baseline have a negative advantage.

The gradient in Eq.~\eqref{grad_integral} can be transformed using these new objects as follows:
\begin{align}
    \nabla_\theta J(\theta) 
    & = \int \left(\sum_{a=1}^A\alpha_a(\p_a)\right) \nabla_\theta \pi_\theta (\p) d\p
    \\ & = \sum_{a=1}^A \int \alpha_a(\p_a) \nabla_\theta \pi_\theta (\p) d\p.
\end{align}

Next, recall that the policy $\pi_\theta(\p)$ is shared among all the agents. For each term $a$ in the sum, we can split the integral over $\p$ into a product of two integrals: over $\p_a$ and $\overline{\p}_a$ (the complement of $\p_a$) and integrate out the complement. This leads to a marginalized policy for each agent $a$ that only depends on $\p_a$:
\begin{align}
    \pi_\theta(\p_a) = \int \pi_\theta(\p) d\overline{\p}_a \label{marginalization} .
\end{align}

Since we work with a simple Gaussian policy, shown in Eq.~\eqref{gaussian_policy}, that factorizes over all components of $\p$, the marginalization in Eq.~\eqref{marginalization} is trivial.

Note that the latter step is not strictly necessary, and we would obtain a valid gradient estimator even without integrating out the irrelevant variables. However, this step plays an important role towards reducing the estimator variance in the case where the dependence of $R_a$ terms on the components of $\p$ is sparse. 

Now the gradient becomes
\begin{align}
    \nabla_\theta J(\theta) 
    & = \sum_{a=1}^A \int \alpha_a(\p_a) \nabla_\theta \pi_\theta (\p_a) d\p_a.
    \label{grad_estimator}
\end{align}

We shall represent it in the form of an expectation value over the samples produced from some distribution. Instead of sampling $\p\sim \pi_\theta(\p)$, we take samples from a different {\it collection} policy $\pi_{\tilde{\theta}}(\p)$ whose variables are denoted as~$\tilde{\theta}$. The use of importance sampling here results in a PPO-type gradient estimator \cite{schulman2017proximal}, while if the samples were taken from the actual $\pi_\theta(\p)$ distribution then we would obtain a REINFORCE-type estimator \cite{williams1992simple}, also known as the score-function estimator \cite{mohamed2020monte}:
\begin{align}
    \nabla_\theta J(\theta) 
    & = \sum_{a=1}^A \int \alpha_a(\p_a) \frac{\nabla_\theta \pi_\theta(\p_a)}{\pi_{\tilde{\theta}}(\p_a)} \pi_{\tilde{\theta}}(\p_a) d\p_a \label{random_label}
    \\ & = \sum_{a=1}^A \underset{\p_a\sim \pi_{\tilde{\theta}}}{ \mathbb{E}}\left[\alpha_a(\p_a) \frac{\nabla_\theta \pi_\theta(\p_a)}{\pi_{\tilde{\theta}}(\p_a)} \right]. \label{intermediate garadient}
\end{align}

Furthermore, in the case of a factorizable policy distribution considered here, the sampling of $\p_a\sim\pi_{\tilde{\theta}}(\p_a)$ in Eq.~\eqref{intermediate garadient} can be replaced with the sampling of $\p$ from the  un-marginalized distribution $\pi_{\tilde{\theta}}(\p)$.

To simplify the following notations, let us define an importance ratio function
\begin{align}
    \chi_{\theta \tilde{\theta}}(\p_a) = \frac{\pi_\theta(\p_a)}{\pi_{\tilde{\theta}}(\p_a)}.
    \label{ir}
\end{align}
For each agent $a$, it is a product of per-parameter importance ratios taken over the policy parameters touched by that agent. With this notation, the gradient becomes
\begin{align}
    \nabla_\theta J(\theta) 
    & = \sum_{a=1}^A \underset{\p\sim \pi_{\tilde{\theta}}}{ \mathbb{E}}\left[\alpha_a(\p_a) \nabla_\theta \chi_{\theta \tilde{\theta}}(\p_a) \right].
    \label{grad_near_final}
\end{align}

Next, we introduce a sparse matrix $S\in \mathbb{R}^{A\times P}$ whose rows correspond to the terms in $R$ and columns correspond to the components of $\p$. The matrix element $S_{ij}$ is set to $1$ if the term $R_i$ depends on the component $p_j$ and $0$ otherwise (hence $S$ can be viewed as a transpose of the Jacobian sparsity matrix). For example, consider a simple function $R(\p) = R_1(p_1) + R_2(p_2) + R_3(p_1, p_2) + R_4(p_2, p_3)$ with $P=3$ and $A=4$. For such a function, the sparsity matrix takes the form
\begin{align}
    S = 
    \begin{pmatrix}
        1 & 0 & 0 \\
        0 & 1 & 0 \\
        1 & 1 & 0 \\
        0 & 1 & 1 
    \end{pmatrix}.
\end{align}

This matrix provides the most efficient description of the dependence of the terms $R_a$ on the components of $\p$. However, we can pretend that each $R_a$ depends on more components of $\p$, although the dependence might be trivial (i.e. there would really be no dependence) -- it corresponds to replacing 0s with 1s in this matrix. Interestingly, any such replacement would still result in an unbiased gradient estimator (even the case $S=\mathbbm{1}^{A\times P}$, although this would be equivalent to global optimization that does not take any advantage of the sparsity).

With the help of the $S$-matrix, we rewrite the importance ratio from Eq.~\eqref{ir} in the vector form:
\begin{align}
    \vec{\chi}_{\theta \tilde{\theta}}(\p) 
    = \exp(S\times [\ln \pi_\theta(\p)-\ln \pi_{\tilde{\theta}}(\p)]) \in \mathbb{R}^A,
\end{align}
where $[\ln \pi_\theta(\p)-\ln \pi_{\tilde{\theta}}(\p)]$ is itself a vector of dimension $P$ of per-parameter importance ratios. Similarly, we write the advantage $\vec{\alpha}(\p)$ as a vector of dimension $A$ whose components are given by Eq.~\eqref{advantage}.

Using these vectorized notations, the gradient from Eq.~\eqref{grad_near_final} can be more elegantly written as
\begin{align}
    \nabla_\theta J(\theta) 
    & = \underset{\p\sim \pi_{\tilde{\theta}}}{ \mathbb{E}}\left[\vec{\alpha}(\p) \cdot \nabla_\theta \vec{\chi}_{\theta \tilde{\theta}}(\p) \right]
    \\ & = \nabla_\theta \underset{\p\sim \pi_{\tilde{\theta}}}{ \mathbb{E}}\left[\vec{\alpha}(\p) \cdot  \vec{\chi}_{\theta \tilde{\theta}}(\p) \right].
    \label{gradient_elegant}
\end{align}

We obtain an empirical Monte Carlo gradient estimator based on a batch of samples $\p^{(i)}$, $i\in 1,...,B$, by replacing the expectation value $\mathbb{E}[f(\p)]$ in Eq.~\eqref{gradient_elegant} with the sample mean $\hat{\mathbb{E}}[f(\p)]=\frac{1}{B}\sum_{i=1}^B f(\p^{(i)})$, where $f(\p)=\vec{\alpha}(\p) \cdot \vec{\chi}_{\theta \tilde{\theta}}(\p)$.

For any policy distribution $\pi_\theta(\p)$, the gradient $\nabla_\theta$ in Eq.~\eqref{gradient_elegant} can be computed with automatic differentiation using a {\it policy loss} function
\begin{align}
    L_{\rm policy} = - \underset{\p\sim \pi_{\tilde{\theta}}}{ \hat{\mathbb{E}}}\left[\vec{\alpha}(\p) \cdot  \vec{\chi}_{\theta \tilde{\theta}}(\p) \right] \label{loss_policy}.
\end{align}

The collection policy $\pi_{\tilde{\theta}}(\p)$ in Eq.~\eqref{loss_policy} can be arbitrary. In practice, it is beneficial for this policy to not differ significantly from the current policy $\pi_{\theta}(\p)$ (otherwise, the importance ratios will strongly deviate from 1 which will increase the estimator variance and can result in training instabilities \cite{schulman2017proximal}). Hence, we simply use the ``old'' policy with parameters $\tilde{\theta} = \theta - d\theta$ from before the most recent gradient step.

Finally, we train the baseline $b$ concurrently with the policy to follow the expected reward of each agent using the quadratic loss for advantages:
\begin{align}
    L_{\rm baseline} = \underset{\p\sim \pi_{\tilde{\theta}}}{ \hat{\mathbb{E}}} \left[ \Vert\vec{\alpha}(\p)\Vert^2\right]. \label{loss_baseline}
\end{align}

Overall, for the choice of policy parametrization in Eq.~\eqref{gaussian_policy}, we have in total $2P+A$ optimizable parameters: $P$ components of $\boldsymbol{\mu}$, $P$ components of $\boldsymbol{\sigma}$, and $A$ components of the baseline.

\begin{table*}
    \setlength\extrarowheight{4pt}
    \begin{tabular}{|l | c | c |}
    \hline
    \multirow{2}{*}{\bf Hyperparameter} & \multicolumn{2}{c|}{\bf Dataset} \\
    \cline{2-3}
     & Repetition codes & Surface codes \\
    \hline
    Batch size	& 70 & 70 \\
    Number of training epochs & 50 & 220 \\
    Number of policy steps per epoch	& 20 & 20 \\
    Learning rate of Adam \cite{kingma2014adam}	& $0.001$ & $0.001$ \\
    Gradient magnitude clipping	& $0.1$ & $0.1$ \\
    Importance ratio clipping	& $0.15$ & $0.4$ \\
    Value loss function coefficient	& $200$ & $200$ \\
    Entropy regularization coefficient	& 0 & $0.01$ \\
    Initial policy standard deviation	& $0.3$ & $0.3$ \\
    Number of QEC shots per epoch & 37,500 & 3,500 \\
    \hline
    \end{tabular}
    \caption{Hyperparameters used for the repetition code and surface code datasets. The role of these hyperparameters is explained in Appendix~\ref{appendix:RL}.}
    \label{tab:hyperparameters}
\end{table*}

\subsection{Relation to natural evolution strategies}

Finally, we note that the presented algorithm, as an edge-case of reinforcement learning, is also closely related to the Natural Evolution Strategies (NES) family of black-box optimization algorithms \cite{wierstra2014natural}. In the terminology of NES, the reward function in Eq.~\eqref{opt_problem} is equivalent to the {\it fitness function}, the policy distribution in Eq.~\eqref{gaussian_policy} is equivalent to the {\it search distribution}, and the gradient of the performance measure in Eq.~\eqref{gradient_of_J} is equivalent to the {\it search gradient}. The search gradient in NES is typically based on the REINFORCE-type estimators \cite{williams1992simple} modified using the Fisher information matrix, while here we use a PPO-type estimator \cite{schulman2017proximal}, as explained prior to Eq.~\eqref{random_label}.

\section{Training hyperparameters \label{appendix:random}}

The behavior of the learning algorithm is regulated using hyperparameters whose role is summarized below.

{\it Batch size} $B$ controls the sampling noise of the Monte Carlo estimate of the policy gradient which tends towards the true gradient in the $B\to\infty$ limit.

{\it Number of training epochs} controls how many times the data will be collected from the learning environment. The data from the previous epoch is discarded when the new epoch begins.

{\it Number of policy steps per epoch} controls how many times the gradient will be evaluated and applied with a fixed set of training data provided in each epoch. 

{\it Learning rate} is a small multiplier for the gradient to ensure that optimization does not make large steps that can lead to instabilities.

{\it Gradient magnitude clipping} is the value to which the magnitude of each gradient component is clipped to prevent instabilities.

{\it Importance ratio clipping} controls the deviation of the importance ratios in Eq.~\eqref{ir} from 1. This parameter was introduced in the PPO algorithm \cite{schulman2017proximal} to mimic the trust-region optimization. 

{\it Value loss function coefficient} controls the relative strength of $L_{\rm baseline}$ loss from Eq.~\eqref{loss_baseline} with respect to $L_{\rm policy}$ loss from Eq.~\eqref{loss_policy}.

{\it Entropy regularization coefficient} controls the relative strength of an additional loss term \cite{haarnoja2018soft}
\begin{align}
    L_{\rm entropy} = - \underset{\p\sim \pi_{\tilde{\theta}}}{ \hat{\mathbb{E}}} \left[ H(\pi_\theta(\p)) \right]
\end{align}
that favors stochastic policies by encouraging increasing the policy entropy defined as
\begin{align}
    H(\pi_\theta(\p)) = -\int \pi_\theta(\p) \ln \pi_\theta(\p) d\p.
\end{align}

Additional hyperparameters relevant for our specific optimization problem are the following.

{\it Initial policy standard deviation}	defines the initial width of each component of the Gaussian policy. Since the range of the hyperedge probabilities spans several orders of magnitude, as seen in Fig.~\ref{fig:training_convergence}(b), we chose to work with the logarithms of these probabilities. Hence, the policy standard deviations here are defined in the log-probability space.

{\it Number of QEC shots per epoch} controls the sampling noise in the estimation of the sensor LERs. We choose to use a relatively small number of shots to speed up the decoding at the expense of having noisier LER estimates during the training. 

The values of the hyperparameters used to produce results in Fig.~\ref{fig3} and Fig.~\ref{fig4} are summarized in Table~\ref{tab:hyperparameters}.

\newpage
\bibliographystyle{apsrev_longbib.bst}
\bibliography{references}

\end{document}